\documentclass[prx,twocolumn,floatfix,superscriptaddress,longbibliography]{revtex4-2}
\usepackage[english]{babel}
\addto\extrasenglish{}
\providecommand{\selectlanguage}[1]{}
\makeatletter
\@ifundefined{l@en}{\let\l@en\l@english}{}
\makeatother

\makeatletter
\@ifundefined{l@eng}{\let\l@eng\l@english}{}
\makeatother

\usepackage{graphicx}
\usepackage{amsmath,amssymb,amsthm,mathrsfs,amsfonts,dsfont,mathtools,physics}
\usepackage{epsfig}
\usepackage[utf8]{inputenc}
\usepackage[T1]{fontenc}
\usepackage{braket}
\usepackage[colorlinks=true,citecolor=blue,linkcolor=magenta]{hyperref}
\usepackage{bm}
\usepackage{enumerate}
\usepackage{color}
\usepackage{tikz}
\usetikzlibrary{quantikz2}
\usepackage{appendix}
\usepackage{svg}
\usepackage{float}
\usepackage{etoolbox}

\usepackage{enumitem}
\usepackage[capitalize]{cleveref}
\usepackage[normalem]{ulem}
\usepackage{comment}
\usepackage{xcolor}

\newtheorem{statement}{Statement}

\Crefname{theorem}{Theorem}{Theorems}
\theoremstyle{remark}

\newenvironment{codeavailability}{%
  \par\medskip
  \noindent\textbf{Code Availability.}\quad
}{%
  \par\medskip
}

\newcommand{\sign}{\text{sign}}

\renewcommand{\var}{\mathrm{Var}}

\begin{document}

\title{Quantum-inspired classical simulation through randomized time evolution}

\author{Fredrik Hasselgren}
\email{fredrik.hasselgren@maths.ox.ac.uk}
\affiliation{Mathematical Institute, University of Oxford, Woodstock Road, Oxford OX2 6GG, United Kingdom}

\author{B\'alint Koczor}
\email{balint.koczor@maths.ox.ac.uk}
\affiliation{Mathematical Institute, University of Oxford, Woodstock Road, Oxford OX2 6GG, United Kingdom}

\begin{abstract}
Tensor-network simulations of quantum many-body dynamics are fundamentally limited by entanglement build-up, which leads to exponentially growing computational costs. Furthermore, these classical simulation algorithms are inherently sequential as typically a tensor network representation of the quantum state is updated incrementally at each time step. We build on recently introduced randomised quantum algorithms for time evolution (TE-PAI), and adapt them to the classical simulation context with the purpose enabling massive parallelisation. Our MPS TE-PAI approach achieves exact time evolution on average (unbiased estimator) and proceeds by representing an ensemble of randomized shallow Trotter-variant circuits as tensor networks. As each circuit instance yields a deterministic quantum state (or observable expected value), the only source of randomness is the sampling of circuit variants -- the absence of shot noise therefore yields a reduced estimator variance relative to quantum hardware implementations of TE-PAI. We simulate representative disordered one-dimensional spin-ring Hamiltonians, and numerically observe reductions in the per-sample gate-count by a factor up to $10^3$ relative to Trotterized MPS evolution, yielding orders of magnitude reduction in the  time-to-solution under realistic levels of parallelisation. Finally, we numerically observe that MPS TE-PAI is substantially more robust against severe bond-dimension truncation than product formulas potentially making it useful for the simulation of strongly correlated systems where truncation is necessary in practice. We also demonstrate that the approach can be used naturally in combination with existing time evolution algorithms, effectively extending their time depth via parallelisation.
\end{abstract}

\maketitle

\section{Introduction}

Accurately simulating the time evolution of quantum systems is a crucial enabler of scientific progress.
Classical state-vector simulation algorithms, however, require either exponential space or time as we increase the system size.
On the other hand, sophisticated matrix product state (MPS) tensor-network techniques can indeed represent certain classes of quantum states
efficiently, but they still scale exponentially as the entanglement in the system is increased~\cite{waintal_who_2026}.

Currently, quantum computers are limited both in the number of qubits and in the achievable circuit depths
due to decoherence~\cite{yan_limitations_2025}. 
However, rapid progress in hardware and quantum algorithm design is expected to
soon enable large-scale, high-accuracy, long-time evolution of quantum systems beyond capabilities of state-of-the-art classical approximations.
On the other hand, this motivates the challenge of pushing boundaries of classical simulation methods for
time evolution.

Tensor-network contractions and related “quantum-inspired” classical algorithms \cite{huynh_quantum-inspired_2023, boev_quantum-inspired_2023, jung_quantum-inspired_2023, oh_quantum-inspired_2024} have recently received attention
as a means to improve and extend the currently known toolbox of quantum simulation techniques~\cite{berezutskii_tensor_2025, orus_practical_2014, orus_tensor_2019}. In regimes where entanglement becomes large (for example, generic chaotic real-time dynamics with volume-law entanglement), low-bond-dimension tensor-network representations can lose their efficiency, motivating alternative representations that track the evolution of observables rather than
states \cite{waintal_who_2026}.
A prominent example is Pauli propagation (also called sparse Pauli dynamics), which simulates dynamics by evolving local operators
in the Heisenberg picture as a truncated expansion over Pauli strings, and has recently been developed into
a practical and theoretically well established framework for simulating circuits and many-body spin dynamics
in challenging regimes \cite{rudolph_pauli_2025, begusic_fast_2024, rudolph_simulating_2025}.

We build on recent advances in quantum computing algorithms, and
develop a quantum-inspired classical approach for extending the simulation capabilities of classical computers. 
Our classical tensor-network MPS protocol builds on the quantum algorithm called Time-Evolution by Probabilistic Angle Interpolation (TE-PAI) \cite{kiumi_te-pai_2024} that reduces time-evolution circuit depth by replacing a single deep circuit with a statistical ensemble of shallow random circuits. While tensor-network based simulation is well explored in the literature \cite{orus_tensor_2019}, these techniques are inherently serial, propagating the wavefunction step-by-step via short-time propagators. 
In contrast, our quantum-inspired classical algorithm trades total computational volume for
reduced depth and increased parallelisation: we distribute the computation of randomly generated circuit instances
across independent computational threads, thereby reducing 
the total time-to-solution at the cost of increased aggregate computational volume.

\begin{figure*}
    \centering
    \includegraphics[width=0.8\linewidth]{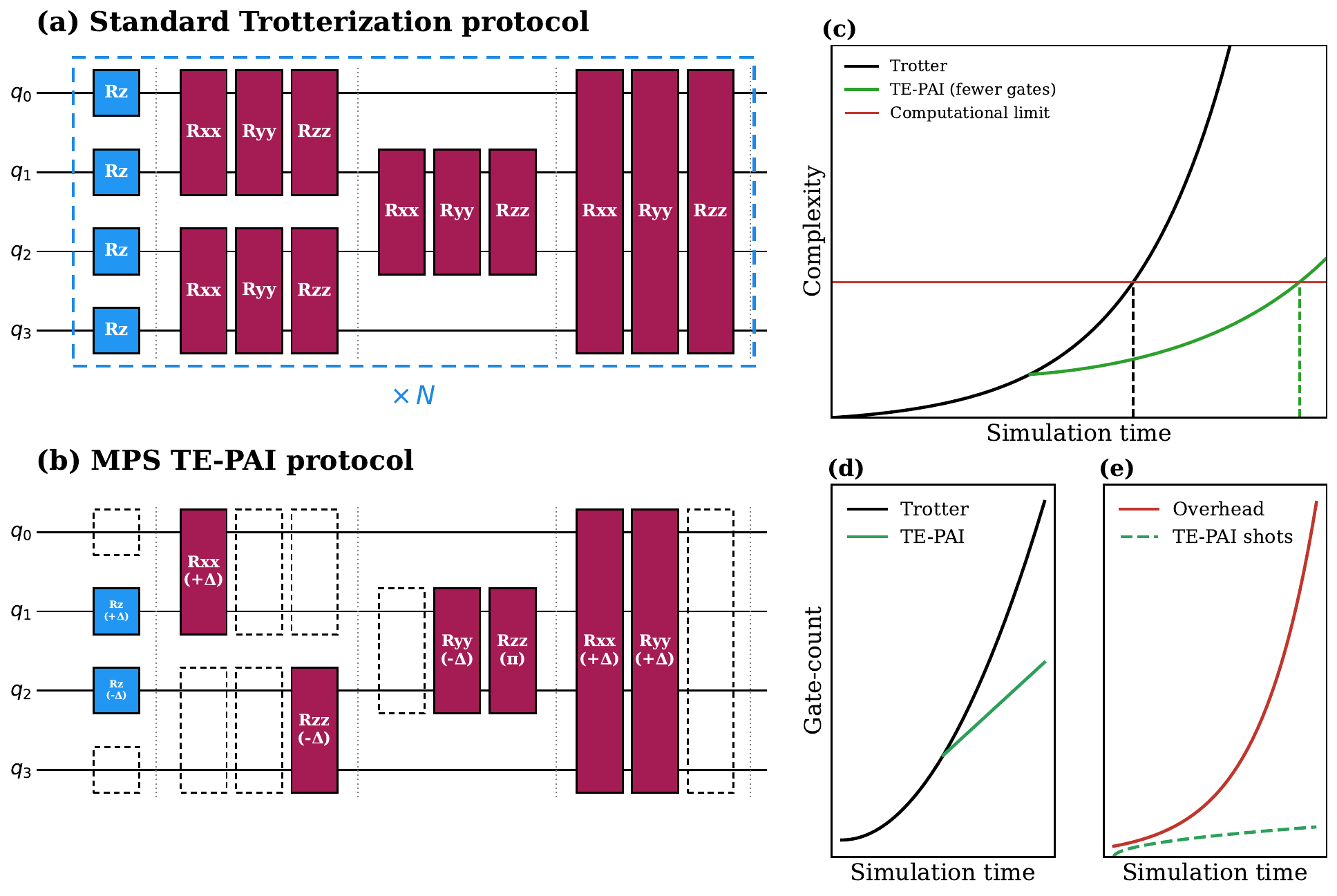}
    \caption{
    	Schematic overview of the MPS TE-PAI protocol and its random Trotter circuit variants. \textbf{(a) \& (b): } circuit diagram for a single Trotter step required to simulate a spin chain Hamiltonian and a randomly generated TE-PAI variant of the circuit. The Trotter rotation gates have their angles probabilistically replaced with $0$ (removal), $\pm\Delta$, or with low probability $\pi$. \textbf{(c-e):} Qualitative plots of the computational advantage of the protocol. \textbf{(c):} Illustrating the exponentially increasing cost of tensor-network simulation 
    	as a function of simulation time and how TE-PAI can extend this simulation time through requiring shallow quantum circuits to be
    	simulated. \textbf{(d):} First-order Trotterization has a gate count that increases quadratically with
    	simulation time (assuming constant error) while TE-PAI has a linearly increasing gate count. \textbf{(e):} Exact (unbiased) simulation with 
    	TE-PAI requires an exponentially increasing number of circuits to be simulated. We confirm in numerical experiments that
    	the true cost is very reasonable, and potentially orders of magnitude below the theoretical bound and present a biased TE-PAI variant that has
    	constant bounded sampling cost.}
    \label{fig:overview}
\end{figure*}

Our \emph{MPS TE-PAI} approach enables unbiased estimation of local observables by averaging an ensemble of shallow randomized Trotter variant circuits that are simulated using efficient tensor network contraction, which we illustrate in \cref{fig:overview}.
In the tensor-network setting, each circuit instance yields a deterministic expectation value, so that shot noise is absent and the sole source of randomness is the sampling of circuit variants; as we detail in \cref{variance}, this leads to a reduced estimator variance.

We perform a range of numerical experiments and demonstrate the following. First, for disordered one-dimensional spin-ring Hamiltonians, we observe substantial practical reductions in per-sample circuit depth relative to deep Trotterization at fixed accuracy,
and we quantify the associated depth–width trade-off under realistic parallelisation assumptions.
Second, we introduce and validate a hybrid simulation strategy that switches from deterministic Trotter evolution to MPS TE-PAI once the bond dimension approaches a truncation threshold,
thereby extending the reachable simulation time under a fixed contraction cost budget.
Third, we demonstrate that TE-PAI is more robust to severe bond-dimension truncation than Trotterization.

This paper is structured as follows. We begin by outlining our cost-metrics in \cref{cost-metrics}. In \cref{sec:product-formulae} we review the main results of the TE-PAI protocol and its connection to standard Trotterization. Next, \cref{variance} derives the expectedly lower variance of our tensor-network approach as well as the special case of biased simulation. In \cref{sec:trotter-baseline} we outline and justify the deep Trotterization protocol used for comparison in this project before relating MPS TE-PAI to other tensor-network techniques in \cref{sec:other-techniques}. The MPS TE-PAI simulations and numerical experiments are outlined in \cref{implementation}. Finally, \cref{sec:discussion} and \cref{sec:conclusion} discuss the conceptual role of MPS TE-PAI, its advantageous regimes, and future directions.

\section{Preliminaries}
\label{sec:prelim}

\subsection{Resource model and cost metrics}
\label{cost-metrics}

\label{sec:resource-model}
To make comparisons between deterministic Trotterized MPS evolution and MPS TE-PAI precise, we adopt an explicit resource model for the classical simulation. Throughout, we consider a one-dimensional $n$-qubit system evolved under a geometrically local Hamiltonian and simulated using MPS, truncating the bond dimension where necessary to maintain computational tractability. Since MPS TE-PAI is a fundamentally parallel protocol that simulates time evolution via an ensemble of random circuit variants we distinguish the following cost metrics:

\begin{enumerate}[leftmargin=*,itemsep=0.2em]
    \item \textbf{Per-circuit cost $C_{circ}$} (runtime for a single circuit instance under serial execution). This serves as a proxy for the floating-point operations (FLOPs) required by a single tensor-network contraction.
    \item \textbf{Total cost $C_{tot}=\sum_i C_{circ, i}$} (aggregate runtime). This is the relevant quantity when parallelisation is unavailable, or when total computational volume is the figure of merit.
    \item \textbf{Time-to-solution $C_{tot}/N_{workers}$} (TTS assuming ideal parallelisation). Under ideal parallelisation with one worker per circuit, the TTS reduces to the contraction time of the deepest single circuit instance. 
\end{enumerate}
The aim of MPS TE-PAI is to attain a lower TTS since each circuit instance satisfies $C_{circ} < C_{circ}^{Trotter}$, yielding a smaller TTS under sufficient parallelisation, albeit at the cost of a potentially larger aggregate computational volume $C_{tot}$.

\paragraph*{Bond-dimension.}
Let $\chi(t)$ denote the maximum bond dimension of an exact MPS at time $t$, and let $\chi_{\max}$ denote the maximum value attained over the full evolution. For a nearest-neighbour two-qubit gate applied to an MPS in mixed-canonical form, the dominant cost of the bond update and SVD re-factorization scales as $\mathcal{O}(\chi^3)$ \cite{orus_practical_2014}. Consequently, for a circuit instance with $\nu$ two-site gates, we model the per-sample cost as
\begin{equation}
    C_{\mathrm{circ}} \;\propto\; \sum_{m=1}^{\nu} \chi_m^{\,3} \;\le\; \nu\,\chi_{\max}^{3},
    \label{eq:Wsample}
\end{equation}
where $\chi_m$ is the bond dimension at the $m$th gate application. This bound makes explicit the two drivers of classical cost: gate count $\nu$ and entanglement-induced bond dimension growth $\chi(t)$.

\paragraph*{Truncation model.}
Bond-dimension truncation is imposed by discarding all but the $\chi \le \chi_{\mathrm{cut}}$, largest singular values at each two-site update, at the cost of a controlled approximation error. In our main numerical experiments we use $\chi_{cut}=16$ at which good convergence is observed.

\subsection{TE-PAI}
\label{sec:product-formulae}
Throughout this work we will compare MPS implementations of  TE-PAI to the first-order Trotter-Suzuki decomposition for time-independent Hamiltonian $H = \sum_{k=1}^L c_k \, h_k $:
\begin{equation}
    e^{-iHT} = \left( \prod_{k=1}^{L} e^{-ic_k h_k \frac{T}{N}} \right)^{N}
    + \mathcal{E},
    \label{Trotter-Suzuki}
\end{equation}
where $\mathcal{E}$ is an error term
bounded in operator norm as $\lVert \mathcal{E} \rVert \leq \epsilon$.
Therefore, achieving precision $\epsilon$ using the above first-order formula requires a gate count $\nu$ that is upper bounded as~\cite{childs_theory_2021,kiumi_te-pai_2024}
\begin{equation}
    \nu \leq \frac{1}{2} L T^2 \lVert c \rVert_T^2 \epsilon^{-1},
    \label{gate-count}
\end{equation}
where $\lVert c\rVert_T^2$ is a commutator norm and we defer further details to \cref{ap:pf_gatecount}.

Rather than implementing all gates in the above product formula, TE-PAI randomly chooses a subset of gates to be implemented at a fixed rotation angle $\Delta$ as per the following protocol.
\begin{enumerate}\label{protocol}
    \item Given an input Hamiltonian, we initialize a Trotter circuit $\mathcal{U}$ in \cref{Trotter-Suzuki}
     with $N$ Trotter-steps over total time $T$. We denote the superoperator representation of this product formula  as
     \begin{equation}
     	\mathcal{U} = \prod_{j=1}^{N} \left( \prod_{k=1}^{L} R_k(\theta_{kj}) \right).
     	\label{unbiased-estimator}
     \end{equation}

    \item Generate $N_s$ random circuit variants by randomly replacing continuous rotation gates $R_k(\theta_{kj})$ in $\mathcal{U}$ with a discrete rotation angle drawn from $\{0, \pm\Delta, \pi\}$ according to the probabilities defined in \cref{ap:te-pai_details}.
    
    \item Execute all circuit variants and in post-processing  multiply each measurement outcome with the relevant prefactor and sign as detailed in  \cref{ap:te-pai_details}.
\end{enumerate}

As discussed in \cref{ap:te-pai_details}, the above protocol provides an unbiased estimator for the time evolution operator as we summarise in the following statement~\cite{kiumi_te-pai_2024,koczor_probabilistic_2024}.
\begin{statement}[From Ref.~\cite{kiumi_te-pai_2024}]
\label{thm:statement1}
Replacing continuous-angle rotations in a Trotter circuit with discrete rotation angles drawn randomly as in the TE-PAI protocol detailed around \cref{unbiased-estimator}, yields the TE-PAI estimator 
$\hat{\mathcal{U}}$ as a single random circuit instance. 
This estimator is unbiased, i.e., $\mathbb{E}[\hat{\mathcal{U}}] = \mathcal{U}$, where the expectation is taken over the random circuit ensemble. 

\end{statement}

The classical computational cost of generating $N_s$ such circuits is $\mathcal{O}(NLN_s)$ \cite{kiumi_te-pai_2024}, 
see further details in \cref{ap:te-pai_details}. Remarkably, starting from an asymptotically deep product formula, the above protocol still yields an expected number of gates that is
finite via the limit as
\begin{equation}
\nu_\infty:=\lim_{N\to\infty}\mathbb{E}[\nu]
    =\csc(\Delta)(3-\cos\Delta) \|\bar{c}\|_1 \,T,
    \label{eq:tepai-gates}
\end{equation}
that scales only linearly with $T$.
The prefactor $\|\bar{c}\|_1$ above is the norm of the Hamiltonian coefficient vector and $\Delta$ is the rotation angle used; please refer to \cref{ap:te-pai_depthstats} for further details. 

The aim of TE-PAI is to estimate the expectation value of a normalized observable $O$, denoted $\langle O \rangle(t)$.
This is achieved by sampling random Trotter circuit variants $\mathcal{U}_{\mathbf{l}}$ such that $\mathbb{E}[\hat{o}_\mathbf{l}] = \langle O \rangle(t)$ yields the true expected value. The only deviation from the ideal expected value is due to finite sampling, as characterised by
the estimator's variance, which grows exponentially in $T$, and determines the sample overhead required to attain a given precision $\epsilon$.
This measurement overhead has been upper-bounded as  $N_s \leq \|g(t)\|_1^2/\epsilon^2$ for TE-PAI~\cite{kiumi_te-pai_2024}, 
where $\lVert g(T)\rVert_1$ is the quasiprobability norm~\cite{cai_quantum_2023}, whose limit for an infinitely-deep Trotter circuit is
\begin{equation}
    \|g(T)\|_1^\infty := \lim_{N\to\infty} \|g(T)\|_1 = \exp \!\left[ 2 \tan \!\left(\frac{\Delta}{2}\right)\|\bar{c}\|_1 \, T \right],
    \label{eq:overhead-main}
\end{equation}
The dependence of $\lVert g\rVert_1^\infty$ on $\Delta$ and the underlying derivations are given in \cref{ap:te-pai_overhead}.

\subsection{Deterministic Expectation Values and Variance Reduction}
\label{variance}

The TE-PAI protocol was originally formulated as a quantum algorithm, whereby expectation values must be estimated from finite-sample projective measurements. In our tensor-network (TN) implementation, however, each sampled circuit instance can be contracted to return the exact expectation value (up to numerical and potential MPS truncation errors) of the target observable for that circuit, so the only remaining source of randomness is the random sampling of circuit configurations. We formalize this distinction in \cref{thm:statement2}, where we derive a tighter variance bound for the TN estimator than that of the quantum implementation; the full derivation is deferred to \cref{ap:tn_variance_proof}.

A further advantage of the MPS implementation is that each sampled circuit $U_l$ applied to the initial state yields an MPS $|\psi_l\rangle = U_l |0\rangle$ that can be stored classically. Together, the stored MPS states and their associated signed coefficients, define a classical representation of the time-evolved state via the signed mixture:
\begin{equation}
    \hat{\rho}(t) = \frac{1}{N_s} \sum_{l=1}^{N_s} \|g(t)\|_1 \, \mathrm{sign}(g_l) \, |\psi_l\rangle\langle\psi_l|,
\end{equation}
from which the expectation value of any observable $O$ can be extracted in post-processing as $\mathrm{Tr}[O\hat{\rho}(t)]$ without additional circuit simulations. In the quantum setting, by contrast, each observable requires independent measurement repetitions. In practice, this enables the simultaneous estimation of a large number of observables from a single set of $N_s$ circuit samples, as we demonstrate in \cref{numerics}.

\begin{statement}
\label{thm:statement2}
In a tensor-network implementation of TE-PAI, each sampled circuit instance $\mathcal{U}_{\mathbf{l}}$ yields the exact (deterministic) expectation value, which when signed gives $\hat v_{\mathbf{l}}:=\sign(g_{\mathbf l}) \Tr\!\left[O\,\mathcal{U}_{\mathbf{l}}|0\rangle\!\langle 0|\right]$, therefore, the sole source of randomness is the random selection of the circuit index $\mathbf{l}$. The variance decomposes as
\begin{equation}\label{eq:config_var}
    \var[\hat{o}_\mathbf{l}(t)] = \|g (t)\|_1^2\, \var[\hat{v}_\mathbf{l}(t)],
\end{equation}
where $\var[\hat{v}_\mathbf{l}] \leq 1$. Therefore, the number of samples required to guarantee precision $\epsilon$ is scaled by the variance of the underlying circuit configurations. 
\begin{equation}
N_s \le \frac{\mathrm{Var}[\hat{o}_\mathbf{l}]}{\epsilon^2}
     = \frac{\|g(t)\|_1^2 \var[\hat{v}_\mathbf{l}]}{\epsilon^2}.
\end{equation}
Although $\|g(t)\|_1$ increases exponentially in $T$, the configuration variance $\var[\hat v_\mathbf l (t)]$ can decay, thereby decreasing the number of samples required to attain a certain accuracy. 
\end{statement}

The reduction in the variance compared to the theoretical overhead is demonstrated in  \cref{fig:variance}, i.e., we numerically
observe that in practice it is common that $\var [\hat v_\mathbf l]\ll1$. Recent work on quasi-probability decompositions (QDP), of which TE-PAI is an example, has explored using stratified sampling \cite{kotz_breakthroughs_1992, cochran_sampling_1977, lohr_sampling_2022} as a technique to reduce the variance \cite{dai_stratified_2026}. The benefits are especially prominent in the case of zero shot noise, which is precisely the context for our tensor-network implementation, where reductions as high as $60-80\%$ have been reported \cite{dai_stratified_2026}. 

As we detailed above, TE-PAI circuits are generated by making a choice between 3 gate-variants for each rotation gate where the probability of choosing a $\pi$ rotation is small. For each $\pi$-flip the sign of the circuit's expected value is multiplied by $-1$, causing the exponential increase in the observables variance. Instead, we may opt to neglect this third rotation angle variant, leading to an approximation which has a variance of at most 1, but the resulting approximation error grows with $T$, as summarized in the following statement (proven in \cref{ap:no_pi_rotations}).

\begin{statement}
\label{thm:statement3}
Neglecting $\pi$-rotations in TE-PAI circuits strictly reduces the estimator variance at the cost of a bias that increases as a function of $T$. In this case, the overhead is constant bounded $\| g\|_1\leq1$  (see \cref{ap:no_pi_rotations}). 
This approach introduces a per-gate bias that scales as $\mathcal{O}(\Delta^2)$ \cite{koczor_probabilistic_2024}. This variant retains the gate-count reduction of MPS TE-PAI while satisfying $\var[\hat{o}_\mathbf{l}(t)] = \var[\hat{v}_\mathbf{l}(t)]$ and $N_s \leq \var[\hat{v}_\mathbf{l}]/\epsilon^2$ at the cost of an overall bias scaling as $\mathcal{O}(T \Delta^2)$.  
\end{statement}
We finally note that rather than omitting the $\pi$ rotation angle, an optimal strategy would
be to use the approach of \cite{PRXQuantum.5.040352} which introduces a controlled bias while simultaneously
minimises the sampling overhead.

\begin{figure}
	\centering
	\includegraphics[width=\linewidth]{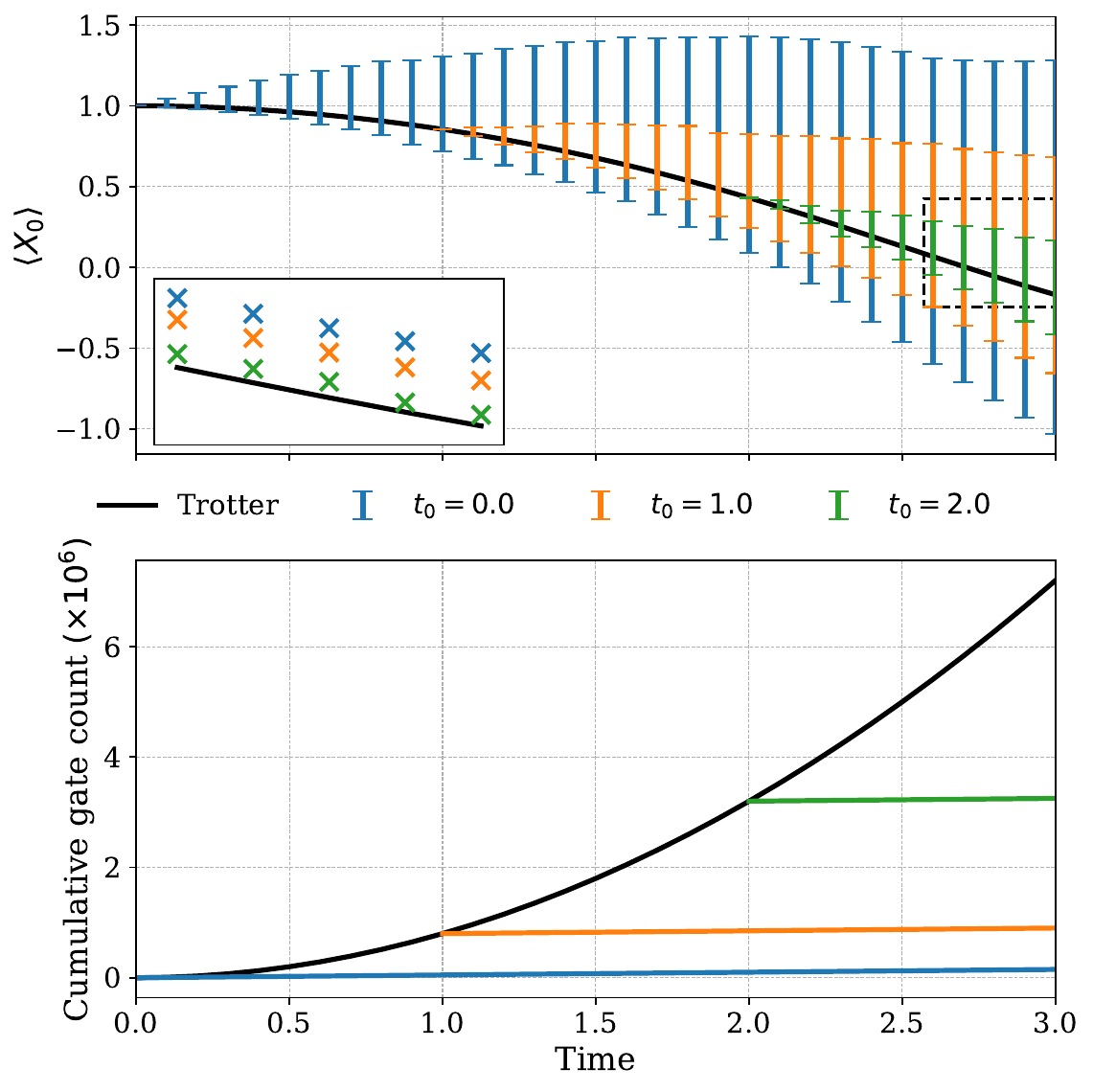}
	\caption{
		Showing Trotterization versus single-sample TE-PAI simulation of $\langle X_0 \rangle$ starting at various times. Upper: Showing the $\langle X_0 \rangle(t)$ for $n=100$ qubits simulated using $\Delta=\pi/2^{10}$. Inset is the single-sample means alongside the Trotterization value for a single sample. Error bars represent the exponential increasing upper bound of the TE-PAI approach~\cite{kiumi_te-pai_2024} which, however, we below demonstrate is pessimistic. Lower: Showing the gate count of the Trotterization versus the TE-PAI protocols starting at various times.}
	\label{fig:manycalc}
\end{figure}

\subsection{Choice of Trotter comparison}
\label{sec:trotter-baseline}

In the numerical comparisons that follow, we take the first-order product formula in \cref{Trotter-Suzuki} as our reference implementation. Specifically, we fix a short reference time $\delta T$ and choose $N_{\delta T}$ such that the first-order formula achieves a desired, sufficiently
small error in the estimated expected value. We then set the number of Trotter steps for desired total simulation time $T$ via
 the quadratic extrapolation
\begin{equation}
    N(T)=N_{\delta T}\left(\frac{T}{\delta T}\right)^2,
    \label{eq:trotter-baseline-scaling}
\end{equation}
so that the gate count obeys the standard constant-accuracy scaling $\nu=\mathcal{O}(T^2)$ used throughout this work \cite{childs_theory_2021, endo_mitigating_2019}. Unless a fixed-$N$ is specified, the numerical Trotterization we used has $\delta T = 0.1$ and $N_{\delta T}=20$.

The parameter $N_{\delta T}$ is used to first fix the prefactors of the scaling according to some desired accuracy, and then the quadratic scaling is used to maintain this accuracy. The Trotter results reported here are therefore not intended to represent state-of-the-art performance in Hamiltonian simulation by product formulae. Rather, they provide a reproducible and unambiguous baseline against which to benchmark the sampling-based TE-PAI construction.

\subsection{Relation to existing tensor-network techniques}
\label{sec:other-techniques}

An advantage of TE-PAI is that it is straightforward to combine with any product-formula-based approach. In particular, our techniques apply directly to, e.g., time-evolving block decimation (TEBD) algorithms for real-time MPS and MPO evolution~\cite{vidal_efficient_2004}, time-dependent density-matrix renormalization group (tDMRG) schemes~\cite{white_real-time_2004}, mixed-state and finite-temperature TEBD algorithms based on matrix-product density operators \cite{zwolak_mixed-state_2004}, and higher-dimensional projected entangled pair state (PEPS) and infinite-PEPS time-evolution methods~\cite{pizorn_time_2011}.

More broadly, MPS TE-PAI fits naturally within the landscape of tensor-network time-evolution methods \cite{paeckel_time-evolution_2019}. At a high level, one can distinguish methods that approximate the propagator's action within a low-dimensional subspace from those that directly approximate the time-evolution operator. The former class includes global Krylov methods~\cite{jose_garcia-ripoll_time_2006,dargel_lanczos_2012,wall_out--equilibrium_2012}, local Krylov schemes~\cite{wall_out--equilibrium_2012,schmitteckert_nonequilibrium_2004,feiguin_time-step_2005,manmana_time_2005,rodriguez_coherent_2006,ronca_time-step_2017}, and one- and two-site time-dependent variational principle (TDVP) algorithms~\cite{haegeman_time-dependent_2011,haegeman_unifying_2016}, which construct the evolved state directly without representing the full propagator. The latter class approximates the time-evolution operator through a time-decomposition analogous to a path-integral representation~\cite{paeckel_time-evolution_2019,keller_feynman_1975}, and includes higher-order TEBD and related MPO $W^{I,II}$ schemes~\cite{white_real-time_2004,zaletel_time-evolving_2015}. For this family, Trotter decompositions can be replaced by TE-PAI, exchanging a portion of the Trotter error for the shallower circuit structure and controlled statistical error of TE-PAI. For Krylov and TDVP-type methods, TE-PAI may be used to construct effective propagators or density matrices entering their subspace projections, potentially transferring the truncation robustness and variance reduction observed here. Systematic investigation of such hybrid constructions is left for future work.

\begin{figure}
	\centering
	\includegraphics[width=\linewidth]{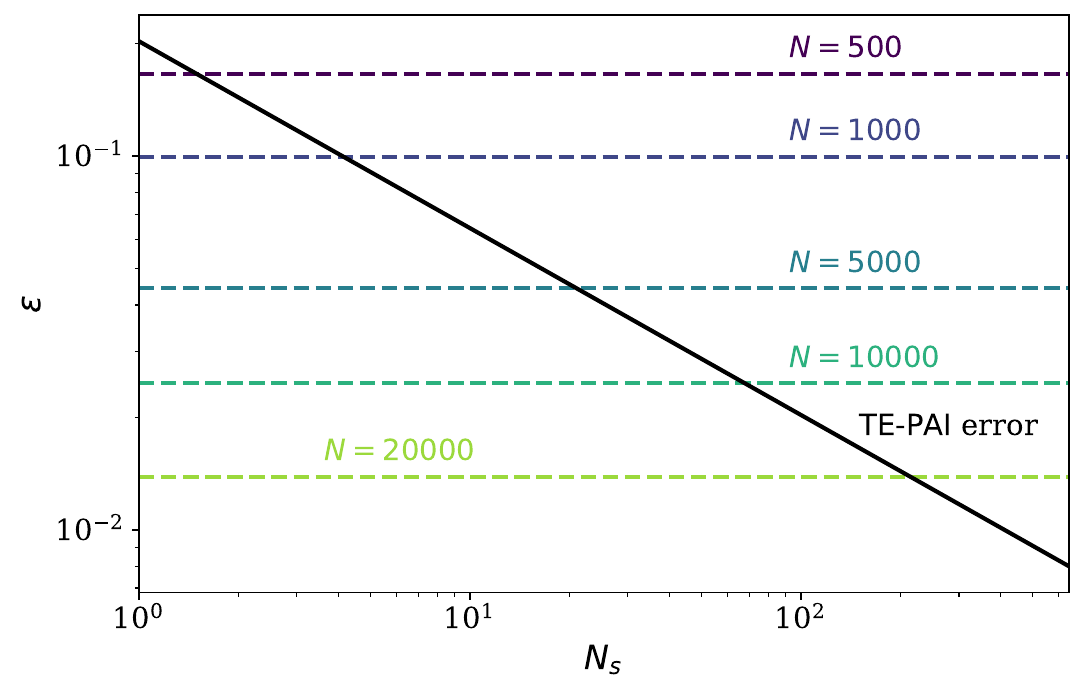}
	\caption{
		Precision $\epsilon$ as the standard deviation in an expected value estimation
		achieved by TE-PAI by averaging $N_s$ circuit samples -- we assume an
		angle $\Delta=\pi/2^{12}$ in a $n=50$ qubit, strongly coupled spin-ring simulation.
		For comparison, the precision is shown that is 
		achieved using $N$ Trotter steps in first-order Trotterization (dashed lines).
	}
	\label{fig:resource-estimation}
\end{figure}

\section{MPS TE-PAI implementation}
\label{numerics}

\label{implementation}
We consider the simple benchmarking the spin-rig Hamiltonian \cref{spinRing}:

\begin{equation}
    H = \sum_{k\,\in \, \text{ring}(n)} \omega_k Z_k + J \, \Vec{\sigma}_k \cdot \Vec{\sigma}_{k+1},
    \label{spinRing}
\end{equation}
letting the coupling strength $J$ serve as a simulation parameter controlling the rate of entanglement growth, and picking parameters $\omega_k$ uniformly randomly within the range $[-1,1]$. As an initial state we chose the state $\lvert +,+,\dots,+,-,+,\dots,+ \rangle$ and measure $\langle X_0 \rangle$ providing a thermalizing observable with which to study the dependence of the estimator variance on simulation time. Numerical simulations of $\langle X_0 \rangle(t)$ were performed comparing Trotterization to MPS TE-PAI with the angle $\Delta$ and number of samples $N_s$ as parameters.

Let us now summarise the primary advantages of MPS TE-PAI which we will benchmark and demonstrate in the rest of this section.

\paragraph*{Circuit parallelisation.}
The MPS TE-PAI protocol is embarrassingly parallel with its circuits being executed independently and their results being cheaply post-processed. This means that computational cost can be traded for parallelized width by spreading the protocol samples across threads, computers, or clusters depending on the system size. This is the classical analogue of the quantum advantage of TE-PAI, in which the per-circuit gate-count is lower than for alternative product formulae and therefore leads to reduced coherent quantum resource requirements.

\paragraph*{Gate-count reduction.}
A further advantage of TE-PAI is its lower per-circuit gate-count relative to Trotterization. Since MPS contraction cost scales linearly with gate-count~\cite{orus_tensor_2019}, and the gate-count of a first-order Trotter circuit grows linearly with the number of steps, this places a further computational constraint on Trotterization which requires deep circuits with gate count bounded by $\nu \propto T^2 \epsilon^{-1}$. In contrast, a TE-PAI circuit requires in expectation a gate-count of $ \nu_\infty =\
(\Delta)(3-\cos\Delta) \|\bar{c}\|_1 \,T$, yielding circuits that are quadratically shallower. The depth-gap clearly grows with increasing precision and total simulation time.

In \cref{fig:manycalc} a $n=100$ qubit simulation demonstrates that the gate-count disparity grows with TE-PAI simulation time. This, however, comes at the cost  of an increased variance due to the exponential increase in the number of samples with simulation time seen in \cref{eq:overhead-main}. As seen in \cref{fig:manycalc}, earlier TE-PAI start times yield quadratically larger gate-count reductions at the expense of a growing variance overhead (error bars), resulting in less accurate sample means. However, as we show in subsequent numerical experiments, this overhead is pessimistic and does not reflect the actual variance growth over time of the tensor-networks.

\begin{figure*}
	\centering
	\includegraphics[width=1.0\linewidth]{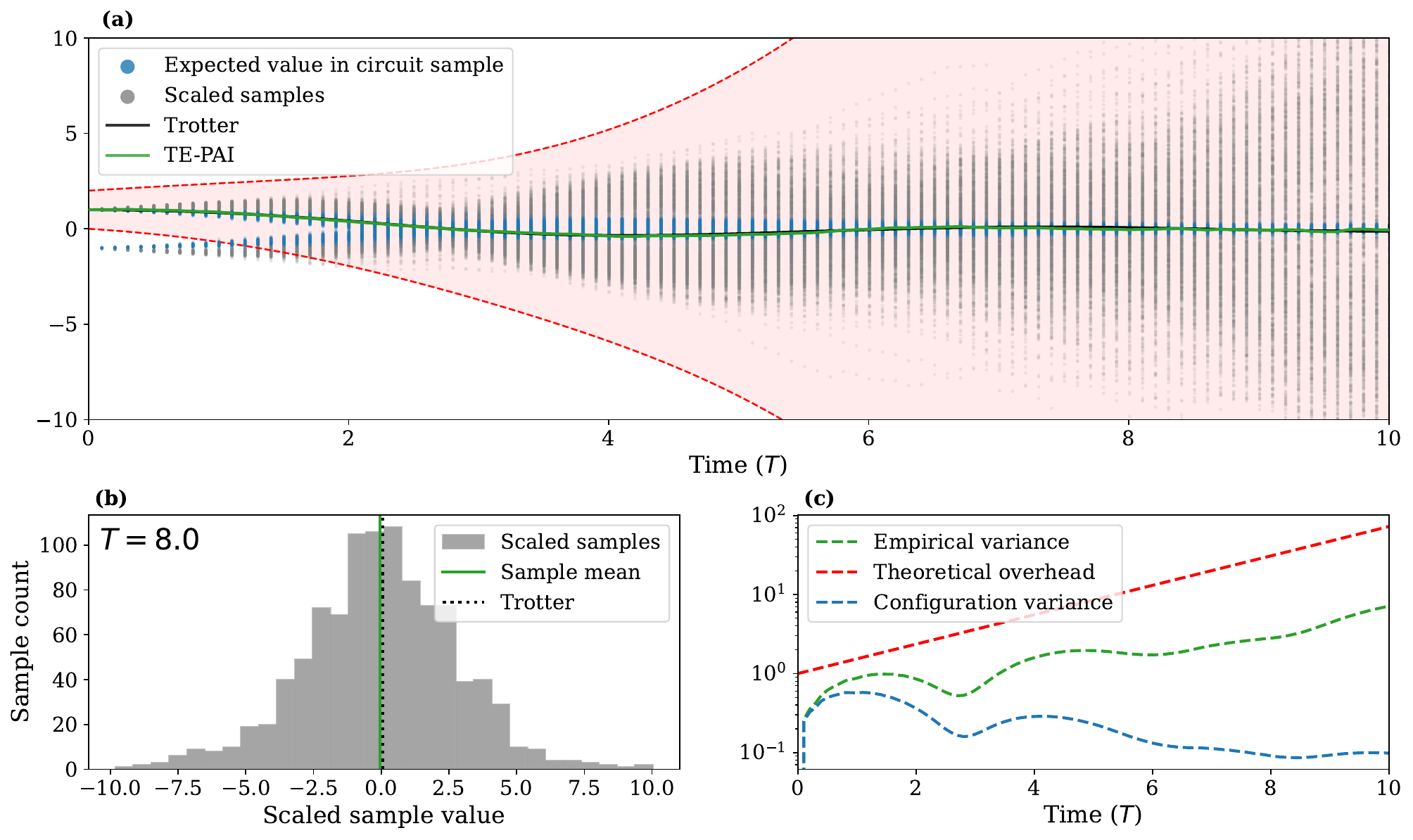}
	\caption{
		Simulating a $n=20$ qubit spin system with $J=0.1$ for up to $T=10$ using $N_s=10^3$ samples of TE-PAI with $\Delta = \pi/2^7$.
		\textbf{(a):} Time-evolved expected values obtained via TE-PAI (green solid) closely match expected values obtained from a deep Trotter circuit (black solid) -- however, each TE-PAI circuit is 3 orders of magnitude shallower.
		Expected values from the individual TE-PAI circuit samples are shown (blue dots) before scaling and after scaling with $\|g (t)\|_1$ (grey dots).
		Also plotted is the theoretical overhead of TE-PAI alongside the actual measurement outcomes of the individual samples. The scaled samples are shown in grey and the raw signed values are shown in blue.
		\textbf{(b):} Estimated variance (green dashed) of the TE-PAI estimator is well below its theoretical upper bound of  $\|g (t)\|_1^2$ (red dashed),
		as the estimated circuit configuration variance $\var[\hat{v}_\mathbf{l}(t)]$ in \cref{eq:config_var} is well below its upper bound of one.
		\textbf{(c):} Histogram of the distribution of the TE-PAI samples in \textbf{(d)} (grey dots) at $T=8$ along with its sample mean (green solid) that closely matches the expected value obtained from Trotterization (black dashed).}

\label{fig:variance}
\end{figure*}

\subsection{Reducing TTS via parallelisation}

In this section we quantify the resources required for TE-PAI to attain a desired precision $\epsilon$ and compare against first-order Trotterization. Specifically, we consider a strongly coupled spin-ring Hamiltonian with $J=2$ for $n=50$ qubits for an evolution time of $T=2.5$. \cref{fig:resource-estimation} (black solid) shows the precision $\epsilon$, measured as the standard deviation of the TE-PAI estimator, as a function of the number $N_s$ of samples. The rotation angle in TE-PAI was chosen as $\Delta=\pi/2^{12}$ which then yields an average $\nu = 2.31\times10^4$ number of gates per circuit sample.

We compare this to first-order Trotterization. Specifically, \cref{fig:resource-estimation} (dashed horizontal lines) shows the precision $\epsilon$ achieved by first-order Trotterization with $N$ Trotter steps, where $\epsilon$ is estimated as the absolute error relative in expected value relative to a deep reference circuit with $N=10^5$ steps (which we use as an approximation of the exact time evolution).  The total number of gates in the Trotterization approach is $200 N$; the Trotter circuits therefore require $4\times$ and $173\times$ more gates than a single TE-PAI sample at $N=5\times10^2$ and $N=2\times10^4$ respectively.

However, $N_s$ independent TE-PAI circuit samples must be simulated. For example, achieving the same precision $\epsilon$ as Trotterization at $N=2\times10^4$, one requires $N_s\approx200$ samples. Therefore, if all TE-PAI circuits can be simulated simultaneously, e.g., by assuming $\approx200$ parallel computational threads,  TE-PAI can deliver the time to solution $173$ times faster.
In the sequential limit, where only a single thread is available, the TTS is approximately $1.15 \times$ that of Trotterization. Therefore, even parallelising over 2 threads yields a faster time to solution with TE-PAI.

\subsection{TE-PAI variance growth \label{sec:var_growth}}
The theoretical upper bound of the number of samples required for TE-PAI in \cref{variance}  has been known to be loose~\cite{kiumi_te-pai_2024}. However, as established in \cref{thm:statement2}, the variance of MPS TE-PAI is further reduced by the absence of shot noise. This variance reduction is demonstrated clearly in \cref{fig:variance} (middle, green vs. red dashed lines) in which the variance of the estimated expectation value is approximately an order of magnitude below its theoretical upper bound at the maximum simulation time of $T=10$. As we established in \cref{eq:config_var}, the ratio of the theoretical upper bound of $\|g (t)\|_1^2$ in \cref{fig:variance} (middle, red dashed) and the true variance in  \cref{fig:variance} (middle, green dashed) is the configuration variance $\var[\hat{v}_\mathbf{l}(t)]$ which we plot in \cref{fig:variance} (middle, blue dashed) and confirm that indeed $\var[\hat{v}_\mathbf{l}(t)] \ll 1$.

\begin{figure}
	\centering
	\includegraphics[width=\linewidth]{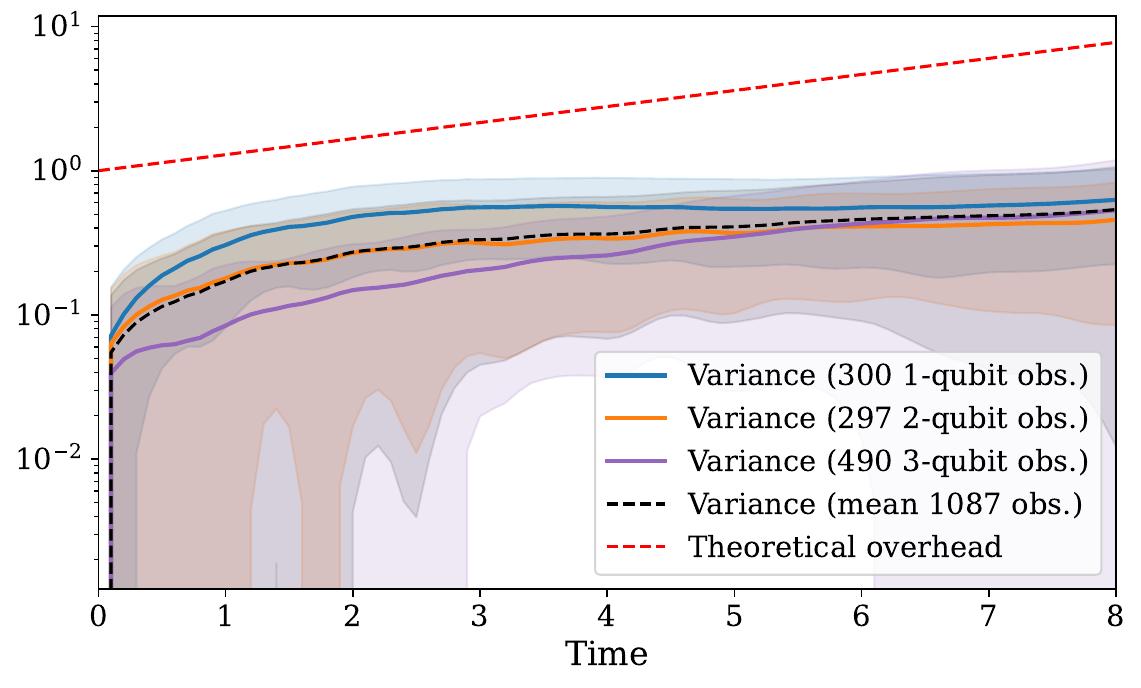}
	\caption{
		Simulating a $n=100$ qubit system with TE-PAI using $N_s=10^3$ samples and a rotation angle of $\Delta=\pi/2^{12}$.
		The average variance of the TE-PAI estimator for all single-qubit Pauli expected values (blue, solid) appears to grow more
		slowly than the exponential growth of the theoretical upper bound (red, dashed).
		Increasing the locality of the Pauli string appears to increase the growth rate of the variance as a function of $T$,
		ultimately approaching the exponential growth rate of the theoretical upper bound.
		Furthermore, variances of all observables are indeed well below the theoretical upper bound, and the variance
		of some observables is orders of magnitude below the bound (shaded areas).
	}
	\label{fig:reduction}
\end{figure}

A lower variance implies fewer samples are required, leading to a substantial reduction in TTS, as demonstrated in \cref{fig:variance} (\textbf{(a)}, green solid line) whereby TE-PAI with $N_s=10^3$ samples accurately reproduces the time-evolved expectation value, compared against a deep Trotter circuit with a total gate count of $\nu_{Trotter}=1.6\times 10^6$ gates in our deep Trotterization protocol. In contrast, each TE-PAI circuit uses an average o $\nu_{TE-PAI}=1422$ gates, a factor of $1125\times$ fewer than the Trotter reference. As such, computing expected values from TE-PAI random circuits in parallel using $10^3$ parallel computational threads reduces the TTS by approximately three orders of magnitude. Additionally, we show the expected values obtained from each individual circuit sample in \cref{fig:variance} (\textbf{(a)}, blue dots) before scaling with $\|g (t)\|_1$ and after scaling in \cref{fig:variance} (\textbf{(a)}, grey dots). A histogram of their distribution is then shown in \cref{fig:variance} \textbf{(b)} along with the sample mean (green solid line) which closely matches the Trotter reference (black dashed line).
¨
We further investigate the looseness of the upper bound and, specifically, its dependence on the locality of the observable. We therefore simulate a larger spin system of $n=100$ qubits for $T=8$ and $J=0.1$, using TE-PAI with $N_s=10^3$ samples and a rotation angle of $\Delta=\pi/2^{10}$. In \cref{fig:reduction} we plot the time evolution of the variance of all 1-qubit Pauli expected values (blue solid line), i.e., X, Y and Z on all qubits, and a set of randomly chosen geometrically 2-local (orange) and 3-local (purple) Pauli strings. For each observable, the error between the expectation values obtained from deep Trotterization and TE-PAI is aggregated. 

Remarkably, the average variance of the TE-PAI estimator for single-qubit Pauli observables in \cref{fig:reduction} (blue, solid) appears to grow more slowly as a function of $T$ than the exponential
in $T$ growth of the theoretical upper bound \cref{fig:reduction} (red, dashed). This is indeed consistent with theoretical results on the reduced overhead of quasiprobability decompositions as PAI gate variants outside of the light cone of the observable do not increase the estimator variance~\cite{tran_locality_2023, jnane_quantum_2024, koczor_probabilistic_2024}.  Increasing the locality of the observables indeed appears to lead to an increased growth rate, approaching the slope of the theoretical upper bound in the linear-log plot in \cref{fig:reduction} (red, dashed). Furthermore, variances of all observables are indeed well below the theoretical upper bound. Finally, let us note that this example also nicely demonstrates the advantage of MPS TE-PAI that classically simulating random circuits enables us to represent and store the time-evolved state classically as a set of MPS states which then allows us to extract any observable efficiently.

\begin{figure*}
	\centering
	\includegraphics[width=1\linewidth]{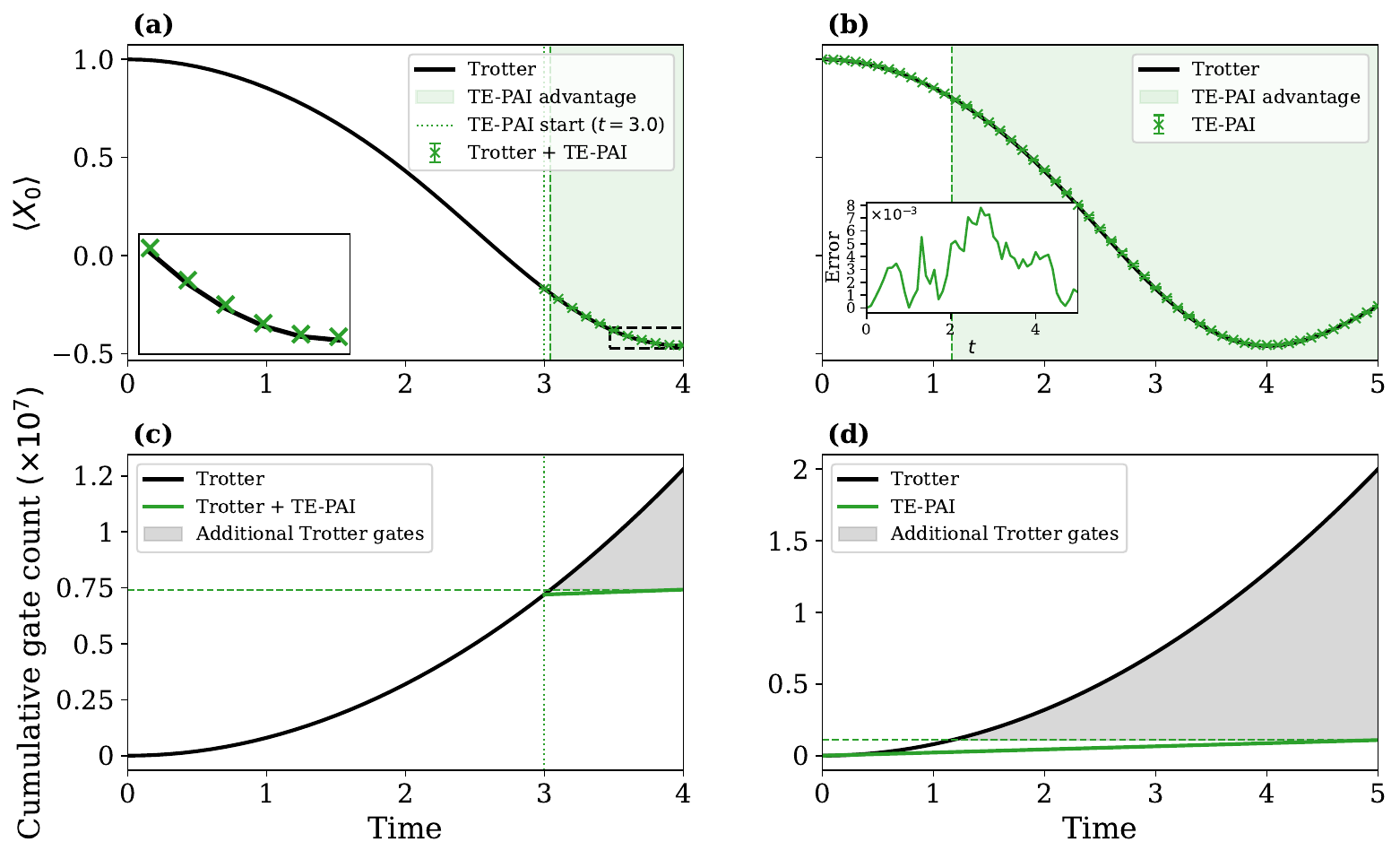}

	\caption{
		Biased low-variance MPS TE-PAI simulating $n=100$ qubits with $J=0.1$ using $N_s=10$ samples with parameter angle $\Delta = \pi/2^{12}$. \textbf{(a):} Shows $\langle X_0\rangle(t)$ simulated with deep Trotterization as well as switching to TE-PAI at $t=3$. The green shaded region is the region of additional simulation reachable for TE-PAI compared to Trotterization at the same computational depth. Inset is a scatter-plot of the last data-points demonstrating the accuracy of the simulation. \textbf{(b):} Compares deep Trotterization to MPS TE-PAI for the entire simulation up to $T=5$, with the advantageous region again shaded green. Inset is the sample mean error of the TE-PAI simulation compared to Trotterization as a function of time. \textbf{(c):} Shows the gate-count growth over time for the deep Trotterization versus the hybrid TE-PAI approach with the gray shaded volume indicating the extra gate depth needed by Trotterization. \textbf{(d):} Shows the same gate-count comparison just for TE-PAI from the start compared to deep Trotterization.
	}
	\label{fig:TrotterThenTEPAI}
\end{figure*}

\section{Application Examples}
\label{D-depth}

\subsubsection{Hybrid TE-PAI}

Let us illustrate MPS TE-PAI in a practical setting. For short evolution times and limited bond dimension, Trotterization is relatively cheap. Therefore, we evolve for an initial period using a deep Trotter circuit, switching to TE-PAI once the bond dimension approaches a truncation threshold. Furthermore, in this example we apply the bounded-variance MPS TE-PAI in \cref{thm:statement3} which has a bias that grows with the simulation time $T$ -- this hybrid approach extends the high-accuracy Trotterization regime at substantially reduced gate-count and sample overhead. Varying the point at which Trotterization is succeeded by TE-PAI enables one to explore the trade-off between increased bias (using TE-PAI for longer increases the bias) and reduced gate count. 

\cref{fig:TrotterThenTEPAI} (\textbf{(a)}, \textbf{(c)}) features a simulation of a $n=100$ qubit spin ring for a total duration of $T=4$ but whereby biased MPS TE-PAI is used for the period between $3 \leq T \leq 4 $ with the small rotation angle of $\Delta=\pi/2^{12}$. The relatively small rotation angle guaranteed a low variance which is further decreased by averaging $N_s=10$ samples, and indeed, the biased TE-PAI simulation \cref{fig:TrotterThenTEPAI} (\textbf{(a)}, green crosses) closely matches the deep Trotter evolution in \cref{fig:TrotterThenTEPAI} (\textbf{(a)}, black solid line). \cref{fig:TrotterThenTEPAI} \textbf{(c)} shows the cumulative gate count for deep Trotterization (black solid), confirming that the biased TE-PAI approach (green solid) requires substantially fewer gates; the additional Trotter gates are indicated by the shaded gray region. Taking gate-count as a proxy for computational depth, then the hybrid TE-PAI approach can simulate through $T=4$ at the same computational cost as Trotterization uses to only slightly exceed  $T=3$. Accordingly, the time-evolution region accessible only to MPS TE-PAI at this computational budget is shaded green in \cref{fig:TrotterThenTEPAI}\textbf{(a)}.

\subsubsection{Full TE-PAI}

We now consider the opposite extreme of the trade-off, in which biased MPS TE-PAI is applied over the full evolution $0 \leq T\leq5$, at the cost of an increased bias. This bias, however, can be controlled by decreasing the rotation angle $\Delta$. \cref{fig:TrotterThenTEPAI} shows a spin-ring simulation of $n=100$ qubits with $N_s=10$ samples of biased MPS TE-PAI at $\Delta=\pi/2^{12}$. Indeed, biased TE-PAI in \cref{fig:TrotterThenTEPAI} (\textbf{(b)}, green crosses) closely matches the expected values obtained from a deep Trotter simulation  in \cref{fig:TrotterThenTEPAI} (\textbf{(b)}, black line).The TE-PAI circuits use 1818
18 times fewer gates than Trotterization; consequently, even sequential evaluation of all circuit samples reduces the TTS, with further proportional reductions available through parallelisation. \cref{fig:TrotterThenTEPAI} \textbf{(d)} highlights that TE-PAI requires substantially fewer gates per circuit (green solid line) than Trotterization (black solid line).

Under the assumption that the bond-dimension truncation threshold is rapidly saturated, gate-count serves as the relevant proxy for TTS under full parallelisation. In \cref{fig:TrotterThenTEPAI} the amount of parallelisation is trivial (realizable on a laptop) and so this comparison comes naturally. From \cref{fig:TrotterThenTEPAI} \textbf{(c)}, the hybrid approach reaches $T=4$ at approximately the same TTS as Trotterization requires to reach $T=3$. From \cref{fig:TrotterThenTEPAI} \textbf{(d)}, the full TE-PAI simulation reaches $T=5$ at a slightly higher TTS than Trotterization requires to reach $T=1$.

\subsubsection{Truncation}

Finally, we demonstrate TE-PAI's relative robustness to approximation errors induced by truncating the MPS bond dimension. Truncation is imposed by retaining only the $\chi$ largest singular values at each bond throughout the tensor network \cite{orus_practical_2014}. Truncation permits simulation to proceed beyond the point at which unconstrained bond-dimension growth would render contraction computationally intractable, at the cost of a bias accumulating with $T$. Since TE-PAI estimates time-evolved expectation values by averaging over an ensemble of randomly generated circuits, this averaging can partially cancel truncation-induced errors across individual circuit simulations.

\begin{figure}[tb]
	\centering
	\includegraphics[width=\linewidth]{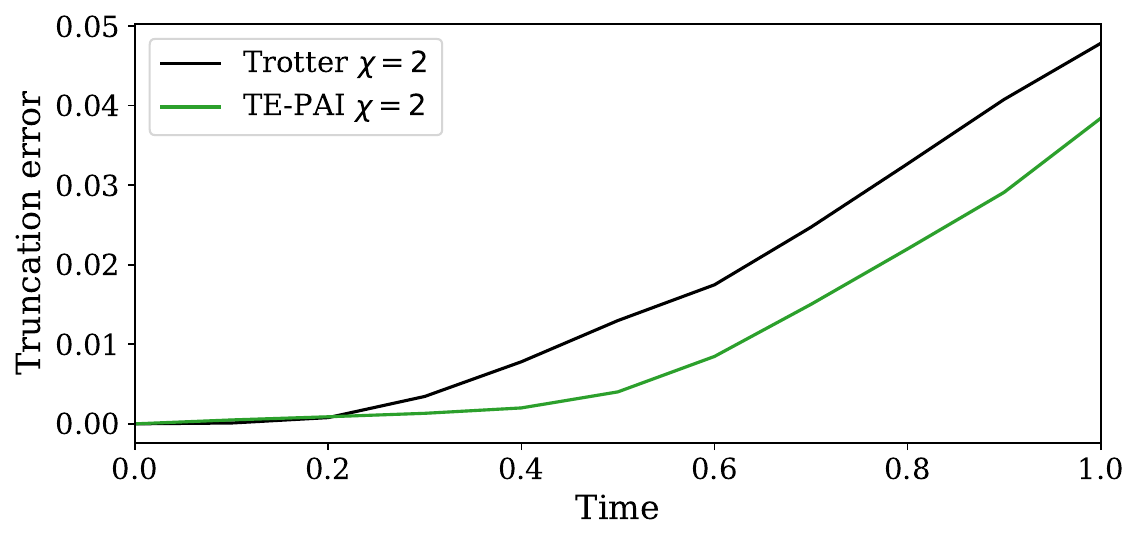}
	\caption{Showing the truncation error of Trotterization versus TE-PAI as a function of time when truncating to $\chi=2$ compared to $\chi=16$.  The truncation error was determined by taking the average of all single qubit observables in the system. The system considered was a strongly coupled $J=1$ system of $n=20$ qubits simulated with $N_s=100$ circuits using $\Delta=\pi/2^{12}$.}
	\label{fig:truncation}
\end{figure}

\cref{fig:truncation} shows the mean truncation error of a simulation of an $n=20$ qubit, strongly coupled $J=1$, spin ring up to $T=1$ by averaging over the error of all single qubit observables. This experiment demonstrates that a heavily truncated, $\chi=2$, TE-PAI achieves a consistently lower truncation error than the equivalently truncated Trotterization. This robustness to bond-dimension truncation error is consistent with partial cancellation of truncation-induced errors across the randomised circuit ensemble.

\section{Discussion}
\label{sec:discussion}

This work considers using the TE-PAI approach from ref.~\cite{kiumi_te-pai_2024} as a quantum-inspired classical simulation technique. We demonstrate advantages of TE-PAI circuits when they are simulated in parallel using MPS techniques. As such, our protocol trades the computational depth of conventional Trotterization for computational width via parallelisation. The following summarises the observed advantages and limitations of the method.

\textit{Gate-count reduction:} In numerical experiments we have demonstrated reductions in gate count of up to three orders of magnitude in TE-PAI circuits relative to Trotterization, e.g., in \cref{fig:variance}. As TE-PAI requires us to average a number of random circuit instances, we have demonstrated that the TTS relative to Trotterization can be reduced by three orders of magnitude assuming a parallelisation over $10^3$ computational threads. On the other end of the trade-off spectrum, we have shown that parallelisation over as few as 2 threads can almost halve the TTS.

\textit{Variance reduction:} While TE-PAI has a variance that grows exponentially with the simulation time $T$, the theoretical exponential bound is known to be loose~\cite{kiumi_te-pai_2024}. Furthermore, MPS TE-PAI is expected to have an even lower variance due to the absence of shot noise as we detail in \cref{thm:statement2} -- this is further elaborated on in ref.~\cite{dai_stratified_2026}. Consistently, numerical experiments confirm that the MPS TE-PAI variance is well below the theoretical bound, e.g., in \cref{fig:reduction} we plotted the average variance over $10^3$ different observables and demonstrated at least an order of magnitude reduction. Furthermore, we observed that the variance of 1-local observables appears to grow sub-exponentially, likely attributable to their light cone remaining localised at short to intermediate evolution times in large systems.

\textit{Robustness to bond-dimension truncation:} Computing expected values from each TE-PAI random circuit sample may require truncating the bond dimension in practice when the aim is to simulate long-time evolution. Truncation, however, introduces an error into the expected value, however, the resulting error is specific to each circuit sample. Averaging over circuit samples is therefore expected to partially cancel truncation errors across trajectories, as we demonstrate numerically.

\textit{Biased low-variance hybrid TE-PAI:} We have also explored a variant of the MPS TE-PAI which neglects the $\pi$-rotation gate variant of the PAI protocol, thereby guaranteeing a constant bounded variance at the cost of a bias that increases with simulation time. For short evolution times from an unentangled initial state, Trotterization remains computationally inexpensive provided the bond dimension stays low. We demonstrated that switching to biased MPS TE-PAI once the bond dimension approaches its truncation threshold substantially extends the simulatable time depth at fixed computational cost.

Another approach we explored in this context is to apply biased TE-PAI for the entire time evolution starting at $T=0$ -- this allowed us to demonstrate a substantial reduction in computational cost compared to Trotterization, see e.g.,  \cref{fig:TrotterThenTEPAI}, whereby the error introduced by the biased TE-PAI approach remained below the level of statistical uncertainty for the entire duration of the time evolution.

\textit{Limitations and future work:} In the present work we have restricted our numerical demonstrations to simple one-dimensional spin-chains~\footnote{While not reported in this work, we have implemented next-to-nearest neighbour and 2D-lattice Hamiltonians and found similarly that TE-PAI indeed outperforms simple Trotterization.}. This motivates further investigation in future work to test whether similar advantages of TE-PAI can be observed in more complex many-body systems, e.g., in PEPS representations. MPS TE-PAI is advantageous only when parallelisation is available; similarly, if the desired expected value needs to be determined to ultra-high precision, then alternative techniques may be superior. Finally, we note that we have used the simple, uniform random sampling approach in TE-PAI, but it has been observed recently that using more advanced stratified sampling can reduce variance further, e.g., by an order of magnitude in ref.~\cite{dai_stratified_2026}. Another possible future improvement dynamically adjusting $\Delta$ and $N_s$ throughout the simulation when the aim is to obtain a series of time-evolved expectation values.

\section{Conclusion}
\label{sec:conclusion}
In this work we have introduced a matrix product state implementation of the algorithm called  Time-Evolution by Probabilistic Angle Interpolation (MPS TE-PAI), as a quantum-inspired classical algorithm for simulating the time evolution of quantum systems. The method replaces a single deep Trotterized circuit with an unbiased ensemble of shallow, randomised circuits that we simulate using MPS techniques. 

Our main technical contributions can be summarised as follows: We adapted the TE-PAI protocol to the tensor-network setting in which the absence of shot noise yields a provably lower estimator variance. We demonstrated that MPS TE-PAI achieves substantial reductions in contraction depth relative to deep Trotterization, whilst maintaining high accuracy even with a low number of samples.
In representative examples, reductions of up to three orders of magnitude were observed at fixed accuracy assuming the ability to process $10^3$ independent TE-PAI circuit simulations in parallel. We further showed that TE-PAI combines naturally with deterministic Trotter evolution: Trotterization is employed while the bond dimension remains low, with a switch to TE-PAI once a truncation threshold is reached thereby extending the simulatable time horizon via parallelisation. Finally, we observed that TE-PAI may be more robust than standard Trotter evolution to severe bond dimension truncation, consistent with partial cancellation of truncation errors across the randomised trajectory ensemble.

In conclusion, as exponential bond-dimension growth fundamentally limits the classical simulation of quantum systems, the present approach extends the reach of classical simulability. TE-PAI is embarrassingly parallel, which we expect is a substantial advantage making MPS TE-PAI well suited to massively parallel hardware architectures such as GPUs.

\begin{codeavailability}
The code used in this work is available online at: https://github.com/fredrikhassel/te-pai-mps, as a package for MPS TE-PAI simulations alongside data from the numerical experiments.
\end{codeavailability}

\begin{acknowledgments}
    The authors would like to thank Jona Erle and Joshua Dai for helpful technical discussions.  FH is thankful for continued support from the Oxford Mathematical Institute Scholarship with Jane Street Graduate Scholarship.
     BK thanks UKRI for the Future Leaders Fellowship Theory to Enable Practical Quantum Advantage (MR/Y015843/1). BK also acknowledges funding from the EPSRC project Robust and Reliable Quantum Computing (RoaRQ, EP/W032635/1). This research was funded in part by UKRI (MR/Y015843/1). For the purpose of Open Access, the author has applied a CC BY public copyright licence to any Author Accepted Manuscript version arising from this submission.
\end{acknowledgments}


\bibliography{references}

\appendix

\section{Deferred derivations and TE-PAI implementation details}
\label{ap:deferred}
\subsection{Derivation of the first-order product-formula gate-count bound}
\label{ap:pf_gatecount}
As shown in~\cite{childs_theory_2021}, the error analysis of product-formula decompositions gives rise to the following bound.
\begin{statement}
\label{statement4}
 The single-step error of the first-order Trotter decomposition satisfies:
 \begin{equation}
    \left\lVert \prod_{k=1}^{L} e^{-ic_k h_k \frac{T}{N}} - e^{-iH \frac{T}{N}} \right\rVert \leq \frac{T^2}{2N^2} \lVert c \rVert_T^2,
\end{equation}
where $\lVert c \rVert_T^2$ is the Trotterization error norm of~\cite{childs_theory_2021},
\begin{equation}
    \lVert c \rVert_T^2 := \sum_{\gamma_1=1}^{L} \left\lVert \sum_{\gamma_2=\gamma_1+1}^{L} \left[ c_{\gamma_2} h_{\gamma_2}, c_{\gamma_1} h_{\gamma_1} \right] \right\rVert.
    \label{error-norm}
\end{equation}
\end{statement}
To obtain the gate-count statement quoted in the main text, we apply the above single-step error bound over $N$ steps and enforce a total simulation error tolerance $\epsilon$. For first-order Trotterization with $L$ exponentials per step the total gate count is $\nu = LN$. Requiring the accumulated error to be at most $\epsilon$ gives
\begin{equation}
    N \geq \frac{T^2}{2}\,\lVert c\rVert_T^2\,\epsilon^{-1},
\end{equation}
and therefore
\begin{equation}
    \nu = LN \leq \frac{1}{2} L T^2 \lVert c \rVert_T^2 \epsilon^{-1},
    \label{ap:eq_gatecount_rederived}
\end{equation}
which is the inequality stated in \cref{gate-count}.
\subsection{Time-dependent Hamiltonians and product-formula notation}
\label{ap:time_dependent_pf}
In the case of a time-dependent Hamiltonian $H(t) = \sum_{k=1}^L c_k(t) \, h_k$ the unitary evolution operator is approximated by a piecewise-constant product formula. Writing $t_j$ for the discrete time grid and assuming that the Hamiltonian remains constant within each time step, one obtains the standard first-order approximation
\begin{equation}
    U(T) \approx \prod_{j=1}^{N} \left( \prod_{k=1}^{L} e^{-ic_k(t_j) h_k \frac{T}{N}} \right),
    \label{trotter-time-dependent}
\end{equation}
whose accuracy increases with $N$, with error bounds discussed in~\cite{childs_theory_2021}. This is the time-dependent analogue of the product formula introduced in~\cref{sec:prelim}.

\subsection{PAI \& TE-PAI}
\label{ap:te-pai_details}

\subsubsection{Summary}

The TE-PAI algorithm \cite{kiumi_te-pai_2024} based on the PAI algorithm
\cite{koczor_probabilistic_2024} approximates the exact time evolution of an
infinitely deep Trotter circuit by sampling from randomly generated shallow
Trotter-variants. We begin by noting that time evolution operators of both
time-independent and time-dependent Hamiltonians employed in both first-order
and higher-order Trotter approximations are approximated by the general form:
\begin{equation}
    U = \prod_{j=1}^{N} \left( \prod_{k=1}^{L} R_k(\theta_{kj}) \right).
    \label{productFormula}
\end{equation}
In this product formula, $R_k(\theta) = e^{-ih_k\theta/2}$ denotes the Pauli
rotation gate applied when evolving a system with Hamiltonian
$H(t) = \sum_{k=1}^L c_k(t) \, h_k$ such that
$\theta_{kj} = 2 \, c_k(t_j) \, \frac{T}{N}$. In the time-independent case
this simplifies to $\theta_{kj} := \theta_k = 2 \, c_k \frac{T}{N}$. Here $T$
is the total evolution time and $N$ is the number of Trotter steps used in the
approximation.
We then use the result from PAI that one can create an unbiased estimator for
a continuous rotation angle using angle-settings of limited resolution. We
randomly replace the rotation angle of $R(\theta)$ with $0$,
$\mathrm{sign}(\theta)\Delta$, or $\pi$ with relative probabilities derived in
\cite{koczor_probabilistic_2024}. We call these gate-settings
\begin{equation}
    A = I, \quad B_{kj} = R_k(\mathrm{sign}(\theta_{kj})\Delta), \quad
    C_k = R_k(\pi).
    \label{settings}
\end{equation}
We assume $\Delta$ is chosen such that $|\theta_{kj}| \leq \Delta \leq \pi$
for all $k \in \{1, \dots, L\}$ and all $j$. In the limit of $N \to \infty$
setting $A$ is almost always chosen and therefore the circuit depth remains
finite and $N$ only affects the complexity of the classical pre-processing
\cite{kiumi_te-pai_2024}.
As proved in \cite{koczor_probabilistic_2024}, the superoperator representation
$\mathcal{R}_k(\theta_{kj})$ of each unitary gate $R_k(\theta_{kj})$ can be
decomposed into
\begin{equation}
\mathcal{R}_k(\theta_{kj}) = \gamma_1(|\theta_{kj}|)\mathcal{A}
+ \gamma_2(|\theta_{kj}|)\mathcal{B}_{kj}
+ \gamma_3(|\theta_{kj}|)\mathcal{C}_k
\end{equation}
such that $\mathcal{A}$, $\mathcal{B}_{kj}$, and $\mathcal{C}_k$ are the
superoperator representations of the unitary gates in \cref{settings}. From
this one can trigonometrically derive the analytic form of the coefficients
$\gamma_l(\theta)$
\begin{align}
    \gamma_1 &= \csc\left( \frac{\Delta}{2} \right) \cos\left( \frac{\theta}{2}
        \right) \sin\left( \frac{\Delta}{2} - \frac{\theta}{2} \right)
        \nonumber \\
    \gamma_2 &= \csc(\Delta) \sin(\theta) \nonumber \\
    \gamma_3 &= -\sec\left( \frac{\Delta}{2} \right) \sin\left(
        \frac{\theta}{2} \right) \sin\left( \frac{\Delta}{2} -
        \frac{\theta}{2} \right),
    \label{gammas}
\end{align}
which determines the selection probability of each gate variant
$\hat{\mathcal{D}}_l \in \{\mathcal{A},\, \mathcal{B}_{kj},\, \mathcal{C}_k\}$
via $p_l(\theta) = \|\gamma_l(\theta)\|_1 / \|\gamma(\theta)\|_1$ for
$l \in \{1,2,3\}$ \cite{kiumi_te-pai_2024}. This leads to the unbiased
estimator for a given continuous-angle gate as
\begin{equation}
\hat{\mathcal{R}}_k(\theta_{kj}) = \|\gamma(|\theta_{kj}|)\|_1 \,
\mathrm{sign}[\gamma_l(|\theta_{kj}|)] \, \hat{\mathcal{D}}_l,
\end{equation}
which was shown in \cite{koczor_probabilistic_2024} to minimize the measurement
overhead.

\subsubsection{Proof of \cref{thm:statement3}}
\label{ap:no_pi_rotations}

We consider the simpler variant in which the $\pi$-rotation setting $C_k$ is
never used. Under the standing assumption $|\theta_{kj}| \leq \Delta$, this
means that each gate is approximated using only the under-rotation
$A = I$ and the over-rotation $B_{kj} = R_k(\mathrm{sign}(\theta_{kj})\Delta)$
from \cref{settings}, respectively. Writing
\begin{equation}
    \lambda_{kj} := \frac{|\theta_{kj}|}{\Delta} \in [0,1],
\end{equation}
the corresponding two-point interpolation of the target superoperator is
\begin{equation}
    \widetilde{\mathcal{R}}_k(\theta_{kj})
    =
    \tilde{\gamma}_1(|\theta_{kj}|)\mathcal{A}
    +
    \tilde{\gamma}_2(|\theta_{kj}|)\mathcal{B}_{kj},
\end{equation}
with coefficients
\begin{equation}
    \tilde{\gamma}_1(|\theta_{kj}|) = 1 - \lambda_{kj},
    \qquad
    \tilde{\gamma}_2(|\theta_{kj}|) = \lambda_{kj}.
\end{equation}
In contrast to the exact PAI decomposition of \cite{koczor_probabilistic_2024},
both coefficients are non-negative, and hence
\begin{equation}
    \|\tilde{\gamma}(|\theta_{kj}|)\|_1
    =
    \tilde{\gamma}_1(|\theta_{kj}|) + \tilde{\gamma}_2(|\theta_{kj}|)
    = 1.
\end{equation}
Thus, at the single-gate level, removing the $\pi$-rotation eliminates the
quasiprobability overhead entirely.
The trade-off is that this two-point formula is no longer exact. Let
$s_{kj} := \mathrm{sign}(\theta_{kj})$. Expanding about $\theta = 0$ gives
\begin{align}
    \mathcal{B}_{kj}
    = \mathcal{R}_k(s_{kj}\Delta)
    &=
    \mathcal{R}_k(0)
    + s_{kj}\Delta\,\mathcal{R}_k'(0) \notag \\
    &\quad
    + \frac{\Delta^2}{2}\mathcal{R}_k''(0)
    + O(\Delta^3), \\
    \mathcal{R}_k(\theta_{kj})
    &=
    \mathcal{R}_k(0)
    + \theta_{kj}\mathcal{R}_k'(0) \notag \\
    &\quad
    + \frac{\theta_{kj}^2}{2}\mathcal{R}_k''(0)
    + O(\Delta^3).
\end{align}
Substituting into the two-point interpolation yields
\begin{equation}
    \widetilde{\mathcal{R}}_k(\theta_{kj})
    =
    \mathcal{R}_k(0)
    + \theta_{kj}\mathcal{R}_k'(0)
    + \frac{|\theta_{kj}|\Delta}{2}\mathcal{R}_k''(0)
    + O(\Delta^3),
\end{equation}
and therefore
\begin{equation}
    \widetilde{\mathcal{R}}_k(\theta_{kj}) - \mathcal{R}_k(\theta_{kj})
    =
    \frac{|\theta_{kj}|(\Delta - |\theta_{kj}|)}{2}\mathcal{R}_k''(0)
    + O(\Delta^3)
    =
    O(\Delta^2).
\end{equation}
Forbidding $\pi$-rotations therefore replaces the exact unbiased decomposition
of \cite{koczor_probabilistic_2024} by a biased two-point interpolation with a
local second-order error.
The remaining conclusions follow directly. Since the negative coefficient
associated with the $\pi$-rotation is absent, all sampling weights are genuine
probabilities and all configuration signs are $+1$. In particular, the variance
overhead originating from the $\ell_1$-norm of the quasiprobability coefficients
is strictly smaller than in the exact PAI construction, where
$\|\gamma(|\theta_{kj}|)\|_1 \geq 1$ \cite{koczor_probabilistic_2024}. The
price paid for this reduction is precisely the $O(\Delta^2)$ bias above.
This extends straightforwardly to the full TE-PAI circuit. If a multi-index
$\mathbf{l} \in \{1,2\}^{LN}$ specifies whether $\mathcal{A}$ or
$\mathcal{B}_{kj}$ is chosen at each gate in \cref{productFormula}, then the
corresponding circuit coefficient factorises as
\begin{equation}
    g_{\mathbf{l}}
    =
    \prod_{j=1}^{N}\prod_{k=1}^{L}
    \tilde{\gamma}_{l_{kj}}(|\theta_{kj}|),
\end{equation}
so that
\begin{equation}
    \|g\|_1
    =
    \sum_{\mathbf{l}} g_{\mathbf{l}}
    =
    \prod_{j=1}^{N}\prod_{k=1}^{L}
    \left(
    \tilde{\gamma}_1(|\theta_{kj}|)
    +
    \tilde{\gamma}_2(|\theta_{kj}|)
    \right)
    = 1.
\end{equation}
Thus, in the no-$\pi$ variant of TE-PAI, the circuit-level overhead also
satisfies $\|g\|_1 = 1$.

\subsection{TE-PAI gate-count scaling and asymptotic depth statistics}
\label{ap:te-pai_depthstats}
Trigonometric and statistical considerations outlined in \cite{kiumi_te-pai_2024} imply that the expected number of gates per circuit sample satisfies
\begin{equation}
    \nu_\infty:=\lim_{N\to\infty}\mathbb{E}[\nu]
    =\csc(\Delta)(3-\cos\Delta) \|\bar{c}\|_1 \,T,
    \label{ap:eq_nu_infty}
\end{equation}
which scales linearly with $T$. This gate-count is bounded from below as $\nu_\infty \geq 2\sqrt{2}\,\|\bar{c}\|_1\, T$, with equality at $\Delta = 2\arctan(1/\sqrt{2}) \approx 0.392\pi$ \cite{kiumi_te-pai_2024}. Furthermore, as proven in \cite{kiumi_te-pai_2024}, in the limit of $N \to \infty$, the number of gates in a given circuit is asymptotically distributed as $\mathcal{N}(\nu_\infty, \sqrt{\nu_\infty})$.
\subsection{Measurement overhead and the $\|g\|_1$ factor}
\label{ap:te-pai_overhead}
The overhead factor is obtained by taking the product of the single-gate quasiprobability norms $\|\gamma_k\|_1$ across all gates in the PAI protocol:
\begin{equation}
    \|g\|_1 = \prod_{j=1}^N \prod_{k=1}^L \left\| \gamma_k(|\theta_{kj}|) \right\|_1.
    \label{overhead-bound}
\end{equation}
Taking the limit $N \to \infty$ and requiring precision $\epsilon$ when estimating the time-evolution of an expectation value, the required number of samples $N_s$ is bounded as
\begin{equation}
    N_s \leq \frac{\left(\|g\|_1\right)^2}{\epsilon^2}.
    \label{eq:Nsbound}
\end{equation}
In the limit $N \to \infty$, the overhead converges to
\begin{equation}
    \|g\|_1^\infty := \lim_{N\to\infty} \|g\|_1 = \exp \!\left[ 2 \tan \!\left(\frac{\Delta}{2}\right)\|\bar{c}\|_1 \, T \right],
    \label{eq:overhead}
\end{equation}
as proven in \cite{kiumi_te-pai_2024}.

\section{Proof of \cref{thm:statement2}}
\label{ap:tn_variance_proof}

Assume the initial state $\mathcal{U}_{\mathrm{circ}} \ket{0}\bra{0}$ is prepared by applying a random circuit as defined in \textbf{Statement~2}, and the goal is to estimate $\Tr[O \,\mathcal{U}_{\mathrm{circ}} \ket{0}\bra{0}]$ for some observable $O$. A circuit instance $\mathcal{U}_{\mathbf{l}}$ is selected at random, where the multi-index $\mathbf{l} \in \{1,2,3\}^{\nu}$ encodes the gate-variant choice from $\{0, \pm\Delta, \pi\}$ at each rotation gate in the Trotter circuit. The index $\mathbf{l}$ is drawn with probability $p(\mathbf{l}) = |g_{\mathbf{l}}|/\|g\|_1$, where $g_{\mathbf{l}}$ is the product of all single-gate coefficients. As derived in \cite{koczor_probabilistic_2024}, each circuit outcome is then weighted by the prefactor $\|g\|_1 \,\sign(g_{\mathbf{l}})$. Here, the sign equals $(-1)$ raised to the number of $\pi$-rotation choices in $\mathbf{l}$. The resulting estimator is
\begin{equation}
    \hat{o}_{\mathbf{l}} = \lVert g \rVert_1 \sign(g_{\mathbf{l}}) \Tr \!\left[O \,\mathcal{U}_{\mathbf{l}} \ket{0} \bra{0}\right].
\end{equation}
The variance follows from the scaling identity $\mathrm{Var}[\lambda \hat{o}] = \lambda^2 \mathrm{Var}[\hat{o}]$, yielding
\begin{equation}
    \mathrm{Var}[\hat{o}_{\mathbf{l}}] = \| g\|_1^2 \,\mathrm{Var}[\hat{v}_{\mathbf{l}}(t)],
\end{equation}
where we defined $\hat{v}_{\mathbf{l}} := \sign(g_{\mathbf{l}}) \Tr[O \,\mathcal{U}_{\mathbf{l}} \ket{0}\bra{0}]$ as the expectation value of a given randomly chosen circuit instance.

\section{Bond dimension growth in MPS TE-PAI}
\label{ap:bond-growth}

The numerical experiments in this work use a truncated bond dimension. If instead an exact MPS representation is desired, one must account for how the bond dimension grows under MPS TE-PAI relative to Trotterization.

\begin{figure*}
    \centering
    \includegraphics[width=\linewidth]{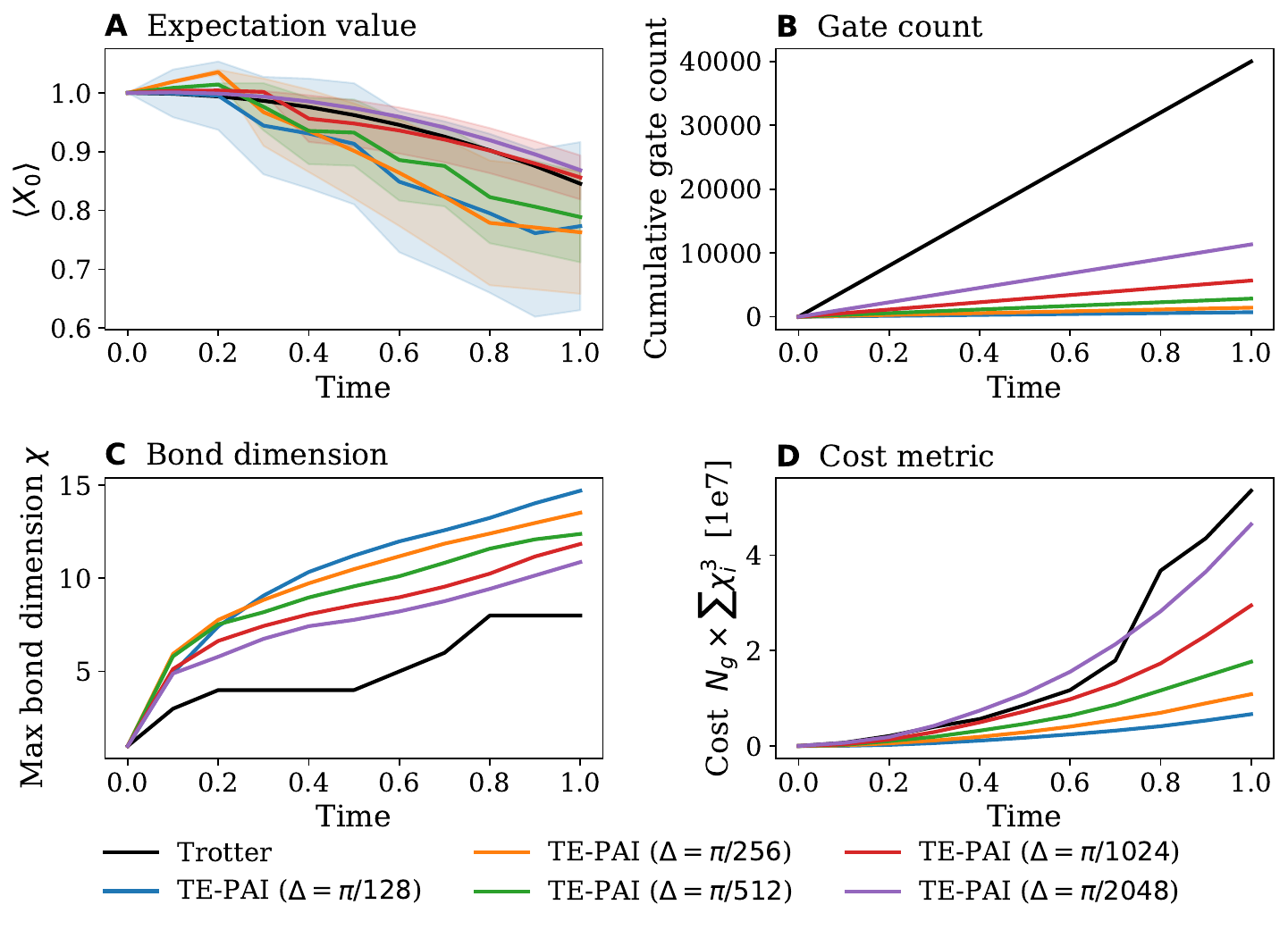}
    \caption{A comparison of $N=1000$ Trotterization and MPS TE-PAI with a range of $\Delta$ values and $N_s=100$ for simulating a system of $n=10$ qubits for $T=1$. \textbf{A:} shows $\langle X_0 \rangle(t)$, \textbf{B:} shows the gate counts over time, \textbf{C:} shows the maximum bond-dimensions needed over time, and \textbf{D:} shows an approximate cost-metric calculated from the bond dimensions and the gate-count over time. }
    \label{fig:deltas}
\end{figure*}

\cref{fig:deltas} compares the bond-dimension growth across TE-PAI simulations with different values of $\Delta$. TE-PAI incurs faster bond-dimension growth than Trotterization when $\Delta$ is large enough that the Trotter gate count is lower. Larger rotation angles increase the rate of entanglement generation, which in turn accelerates bond-dimension growth beyond what the physical evolution requires. As $\Delta$ decreases and the circuit depth grows, the crossover occurs approximately when the TE-PAI and Trotter circuits have comparable gate counts. A proxy for the tensor-network contraction cost can be obtained by summing, over all gates, the cube of the local bond dimension at each site. Under this metric, the gate-count reductions afforded by large-$\Delta$ TE-PAI can outweigh the additional cost of the spurious bond-dimension growth in certain regimes. This advantage vanishes once $\chi$ is sufficiently large that bond-dimension growth dominates the cost metric over gate-count savings; however, freely growing bond dimensions quickly render the tensor-network contraction intractable in practice.

These observations indicate that MPS TE-PAI is most beneficial in the truncated bond-dimension regime, which is standard for large-scale simulations. Once the bond dimension saturates its truncation threshold, the cost metric depends entirely on the gate count, where TE-PAI offers substantial reductions. Moreover, we conjecture that the truncated regime offers an even larger TE-PAI advantage, since truncation errors may partially cancel across the randomised circuit ensemble, yielding lower aggregate bias than in an equivalent Trotter simulation. Applying TE-PAI with a freely growing bond dimension would require a sufficiently large $\Delta$ to prevent the spurious bond-dimension growth from dominating the cost, at which point achieving a gate-count reduction relative to an equivalent Trotter circuit may no longer be feasible.

\begin{figure*}
    \centering
    \includegraphics[width=\linewidth]{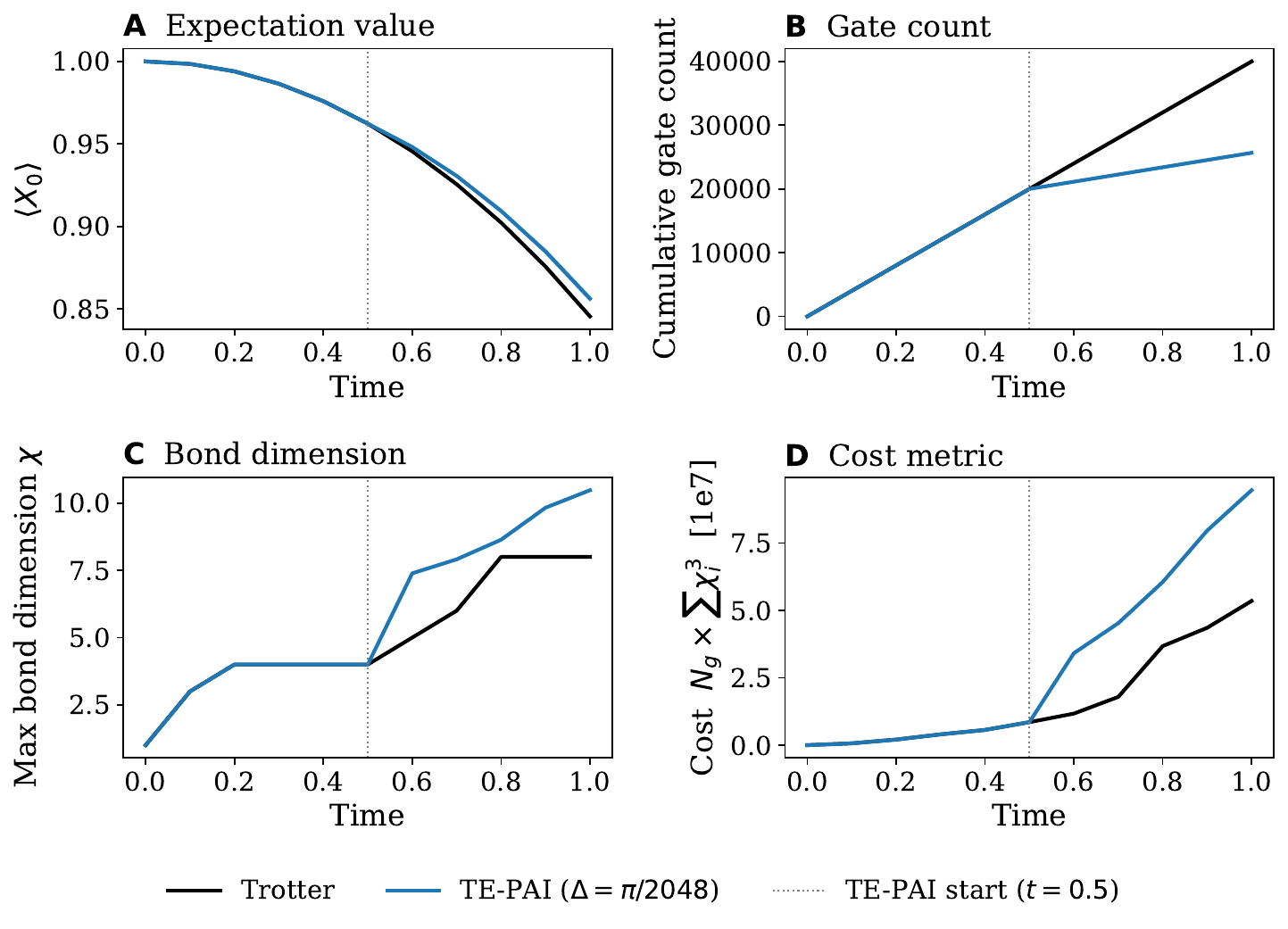}
    \caption{A comparison of $N=1000$ Trotterization and hybrid MPS TE-PAI with $\Delta=\pi/2048$ and $N_s=100$ for simulating a system of $n=10$ qubits for $T=1$. }
    \label{fig:switch}
\end{figure*}

This discrepancy has further ramifications for the hybrid approach. In the hybrid scheme, a deep Trotter circuit is used for the initial period of the simulation to minimise the spurious bond-dimension growth, before switching to TE-PAI. Our analysis of the bond-dimension profiles indicates that the hybrid approach is beneficial only in the truncated bond-dimension regime, as illustrated in \cref{fig:TrotterThenTEPAI}. If the switch to MPS TE-PAI is motivated by the bond dimension having grown large enough to limit further simulability, then the faster bond-dimension growth inherent to TE-PAI renders the switch disadvantageous. \cref{fig:switch} illustrates such a scenario: although the TE-PAI circuits use fewer gates than continued Trotterization, this saving is outweighed by the increased bond-dimension growth. If instead the switch is made at the point where the bond dimension must be truncated to maintain tractability, the gate-count reduction translates directly into a lower TTS.

\end{document}